\documentclass{kluwer}    
\usepackage{epsfig}

\begin{document}                                                                                   

\begin{article}

\begin{opening} 

\title{Modelling the radio to X-ray SED of galaxies} 

\author{Laura \surname{Silva} \email{silva@ts.astro.it}}
\institute{Osservatorio Astronomico di Trieste and SISSA, Trieste, Italy}
\author{Alessandro \surname{Bressan}}
\author{Gian Luigi \surname{Granato}}
\institute{Osservatorio Astronomico di Padova, Padova, Italy}
\author{Pasquale \surname{Panuzzo}}
\institute{SISSA, Trieste, Italy}

\begin{abstract} We present our model to interpret the SED of galaxies. The model for
the UV to sub-mm SED is already well established \cite{silgra98}. We remind here
its main features and show some applications. Recently we have extended the model to the radio
range \cite{bregra01}, and we have started to include the X-ray emission from the stellar component.

\end{abstract}
\end{opening}           

\section{Modelling the UV to sub-mm SED of galaxies}  

\begin{figure} 
\centerline{ \epsfig{file=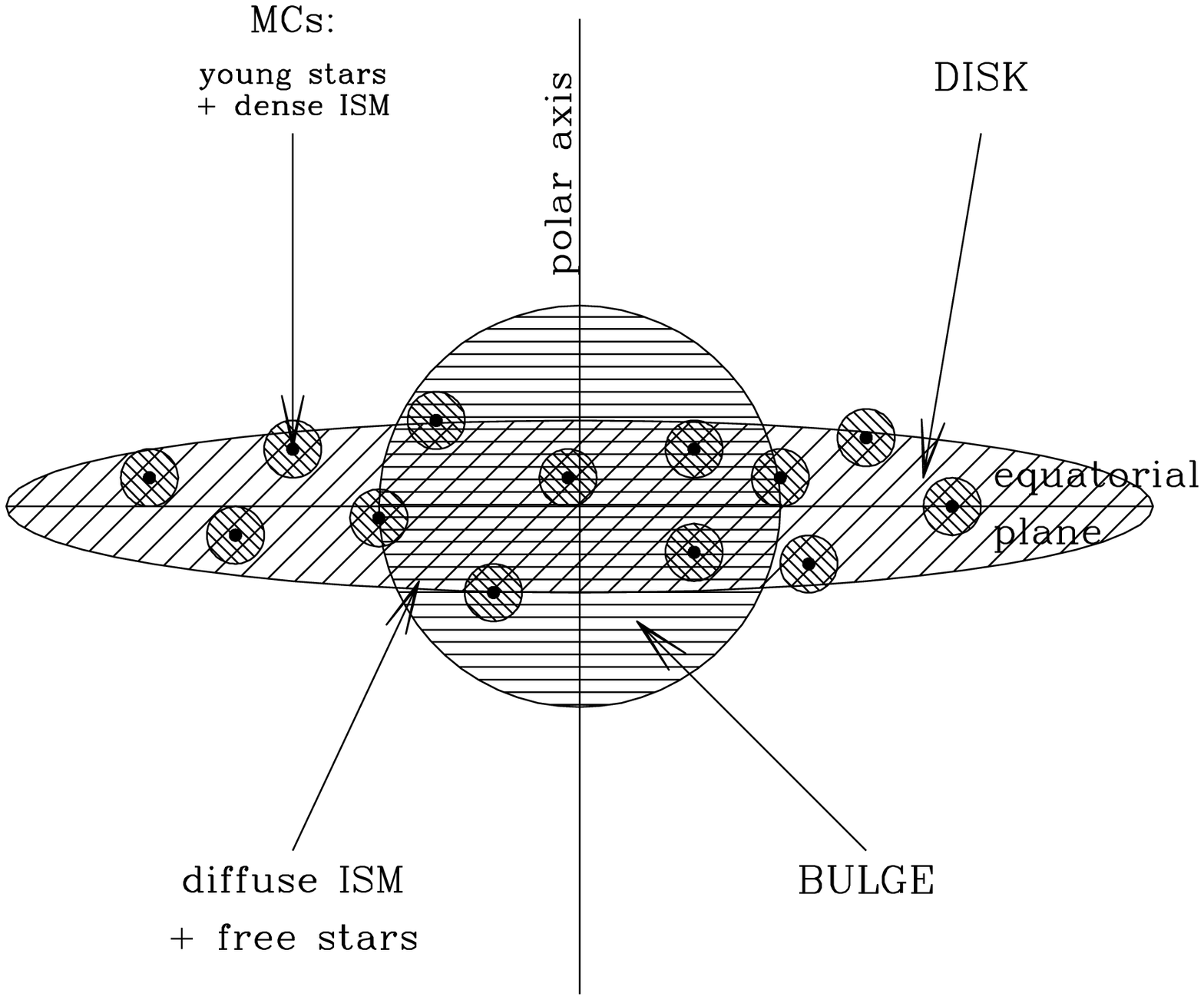,height=6truecm,width=7truecm}     
\epsfig{file=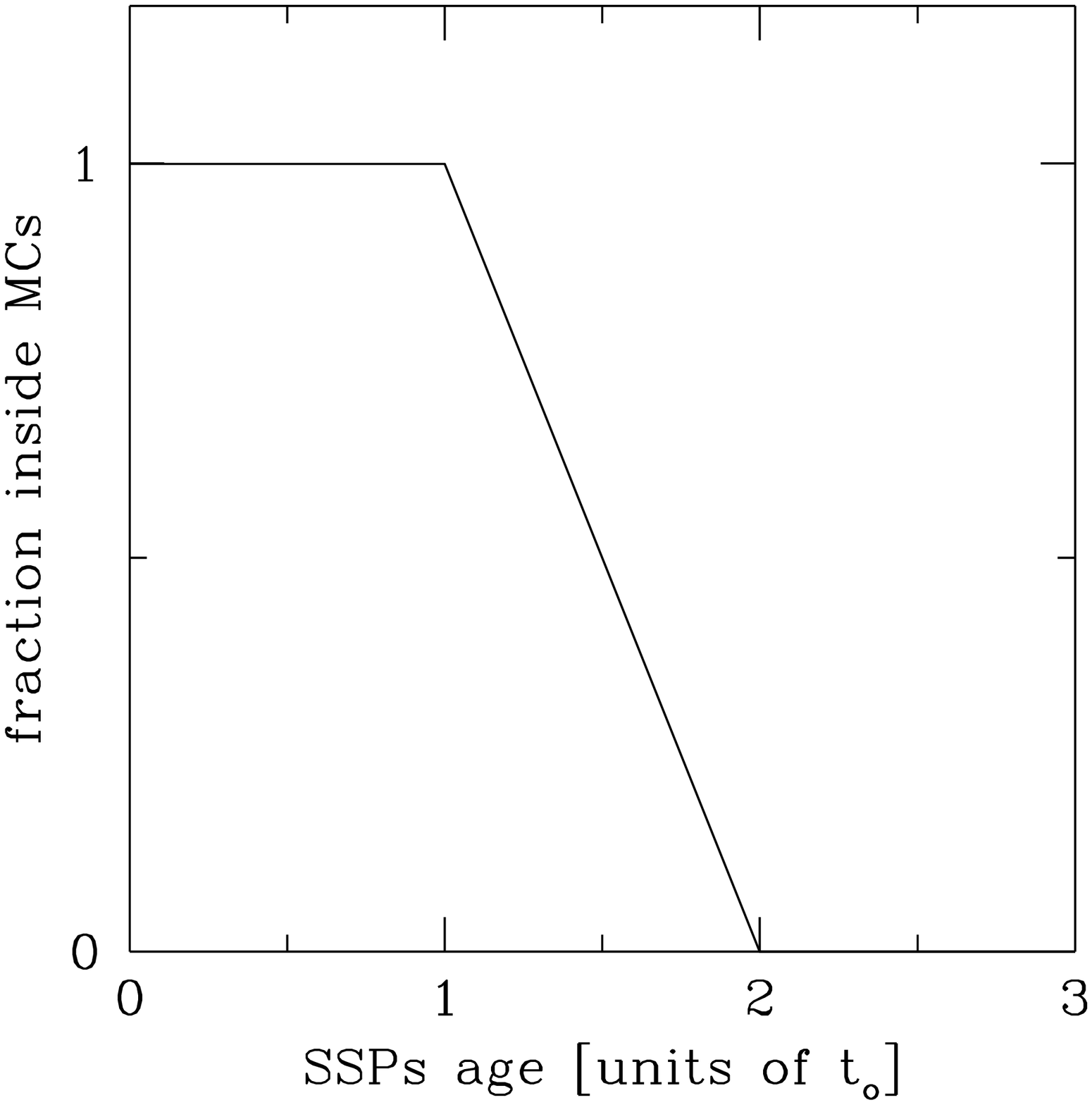,height=4truecm,width=4truecm} }          
\caption[]{{\it Left}: Scheme and components of our model galaxy. {\it Right}: Fraction of SSP enegy inside MCs vs SSP age.}
\label{cartoon2f_mc}
\end{figure}

We have developed a model (GRASIL, {\it http://grana.pd.astro.it}) for the SED of galaxies including dust effects
\cite{silgra98,sil99}. We consider a geometry 
more complex and realistic than previous treatments, and take into account 3 distinct 
environments where stellar radiation interacts with dust (Figure~\ref{cartoon2f_mc}): 
(1) {\bf molecular clouds
(MCs) + young stars} (clumped component), (2) {\bf diffuse ISM (cirrus) + older stars} (diffuse component),
(3) AGB envelopes (included in the SSP library, Bressan et al., 1998). 
Therefore both stars and dust have a {\it realistic 2-phase distribution}.
The association between young stars and MCs is modelled to mimic the birth of stars in MCs and their
gradual escape as they grow old: the fraction of each SSP energy radiated inside MCs is decreased with its
age (Figure~\ref{cartoon2f_mc}). It is the first time that such an {\it age-dependent or selective extinction}
is modelled (see Granato et al., 2000 for interesting applications).    

The dust model consists of normal grains in thermal equilibrium, small grains
fluctuating in temperature and PAH molecules.

\section{Modelling radio emission in star-forming galaxies}  

We have extended our model to the radio.
The thermal radio emission is proportional to the production rate of ionizing photons:

\begin{equation}
\frac{L_{Th}(\nu)}{erg/s/Hz} \simeq \frac{Q(H)}{5.495\times 10^{25}}
\left(\frac{T_e}{10^4\, K}\right)^{0.44}\left(\frac{\nu}{1.4 GHz}\right)^{-0.1}
\end{equation}

The non thermal radio emission is the sum of a dominating term, 
proportional to the supernova rate (calibrated on the
Galactic value) and one due to supernova remnants (based on the $\Sigma-$D relation):

\begin{equation}
\frac{L_{NTh}(\nu)}{erg/s/Hz} = K\times \nu_{SN} \left(\frac{\nu}{1.4}\right)^{-\alpha}+
L_{1.4}^{SNR}\left(\frac{\nu}{1.4}\right)^{-0.5}
\end{equation}

See Bressan (these proceedings) and Bressan et al.\ (2001) for details.

\section{Applications}  

\subsection{Fitting the SEDs}

\begin{figure} 
\vspace{-0.3truecm}
\centerline{ \epsfig{file=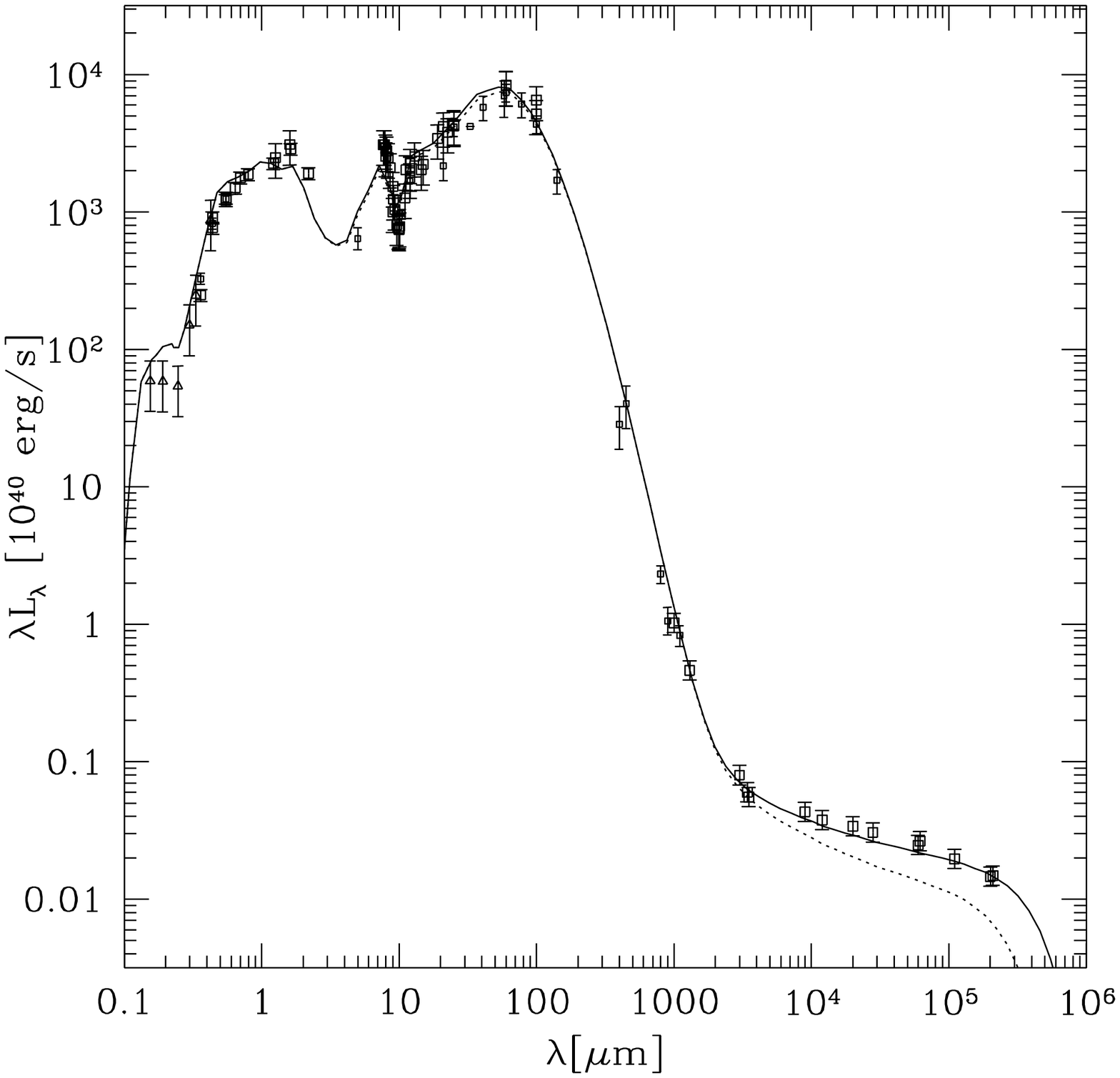,height=5truecm,width=5truecm}  
\epsfig{file=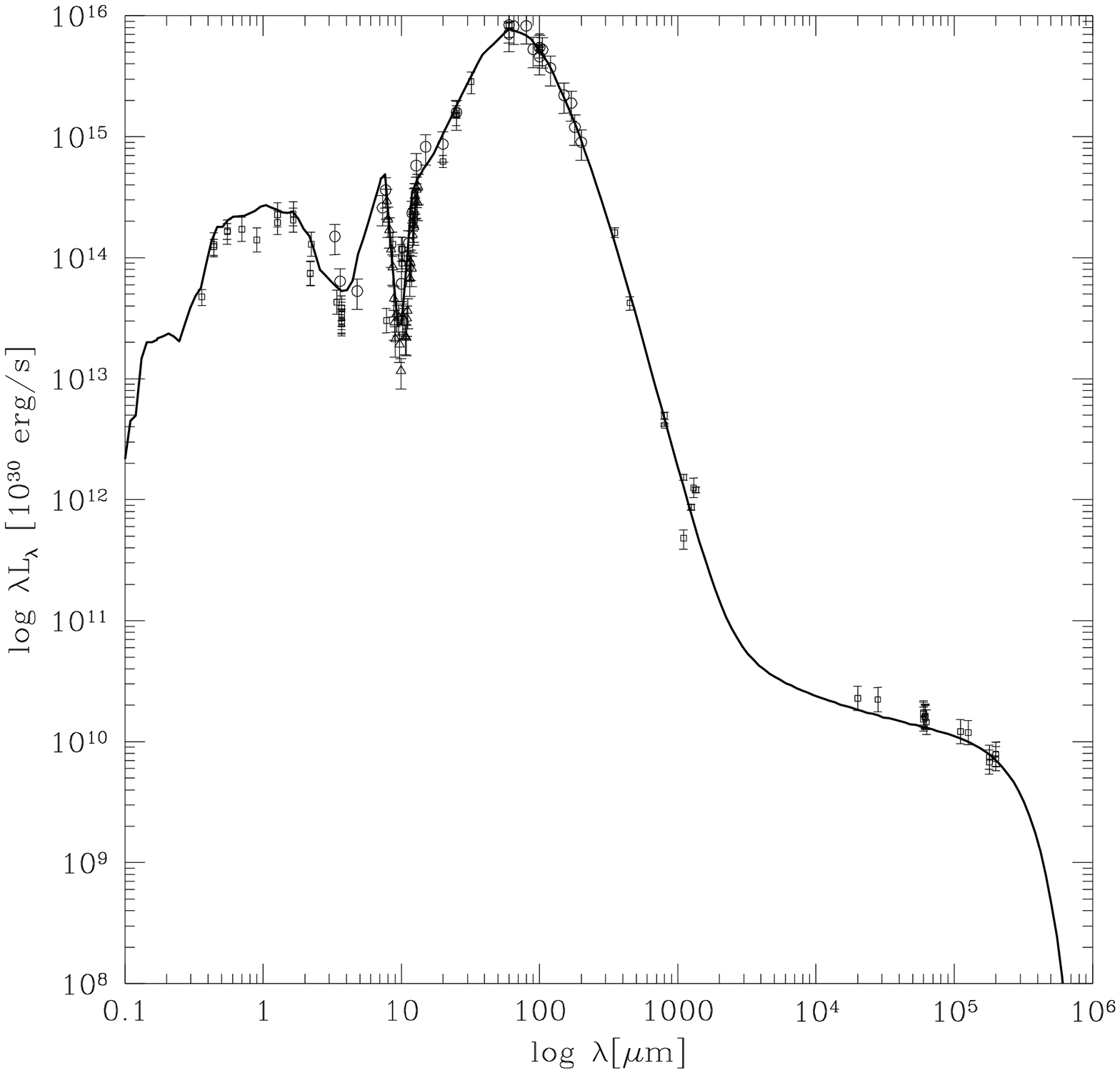,height=5truecm,width=5truecm} }
\caption[]{{\it Left}: Fit to the SED of M82. The starburst age and e-folding time are respectively $25$ Myr and $10$ Myr (continuous line) 
and both $50$ Myr (dotted line). An older starburst is compatible with the UV to sub-mm SED but not with the radio emission. 
A free-free absorption $\tau_{1.4 \mbox{GHz}}=0.2$ has been applied. {\it Right}: Fit to the SED of Arp220.}
\label{m82r}
\end{figure}

As an example, the fit to the starburst galaxy M82 is shown in Figure~\ref{m82r}. Note that
the inclusion of the radio emission potentially (with the adopted calibration) helps in constraining 
the starburst age, due to the different timescales involved in the IR (ongoing SFR, last few Myrs) 
and radio (recent SFR, a few $10^7$ yrs ago) emission.

\subsection{The redshift of sub-mm selected galaxies} 

\begin{figure} 
\vspace{-0.3truecm}
\centerline{ \epsfig{file=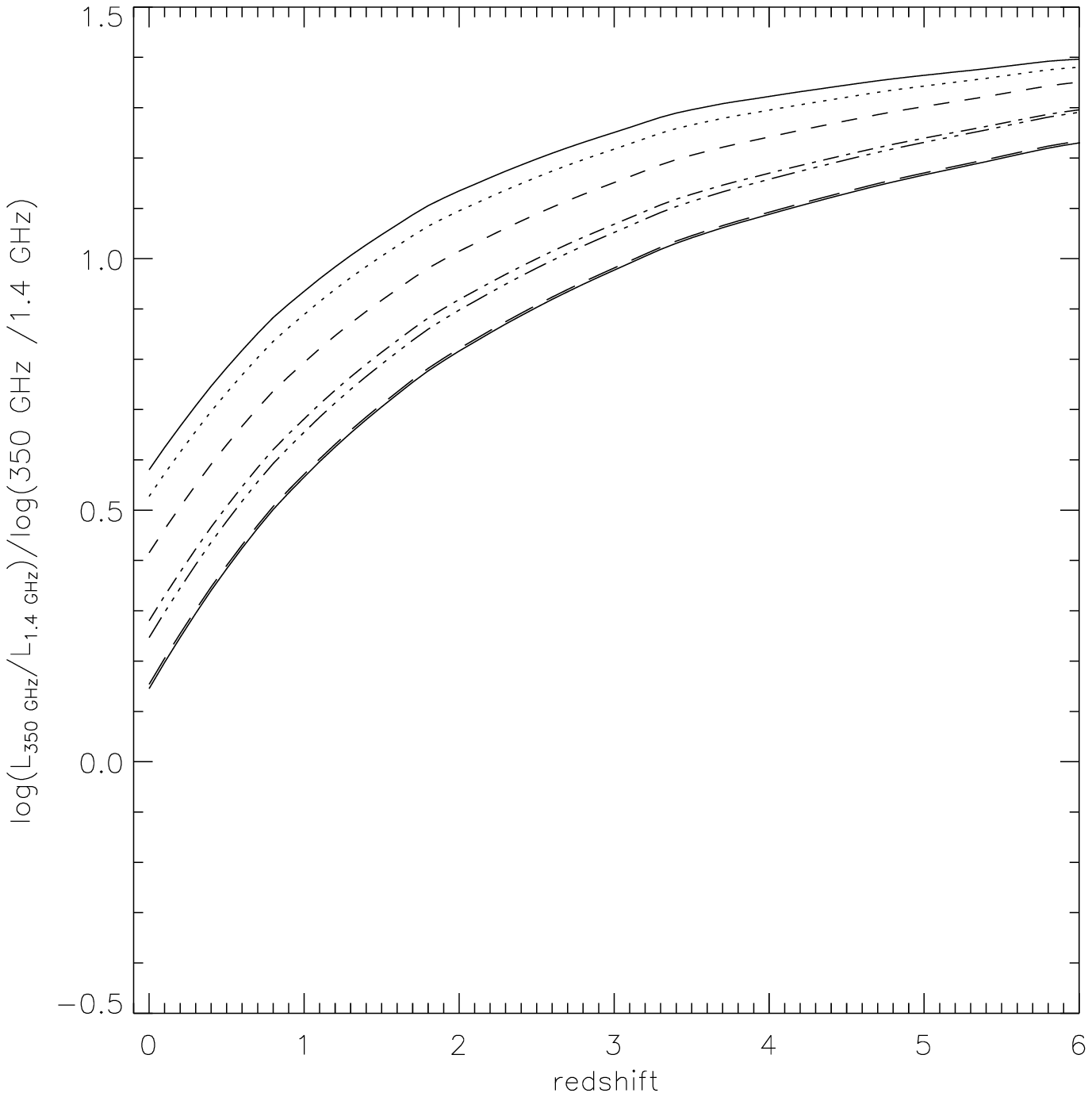, height=5truecm,width=5truecm}  
 \epsfig{file=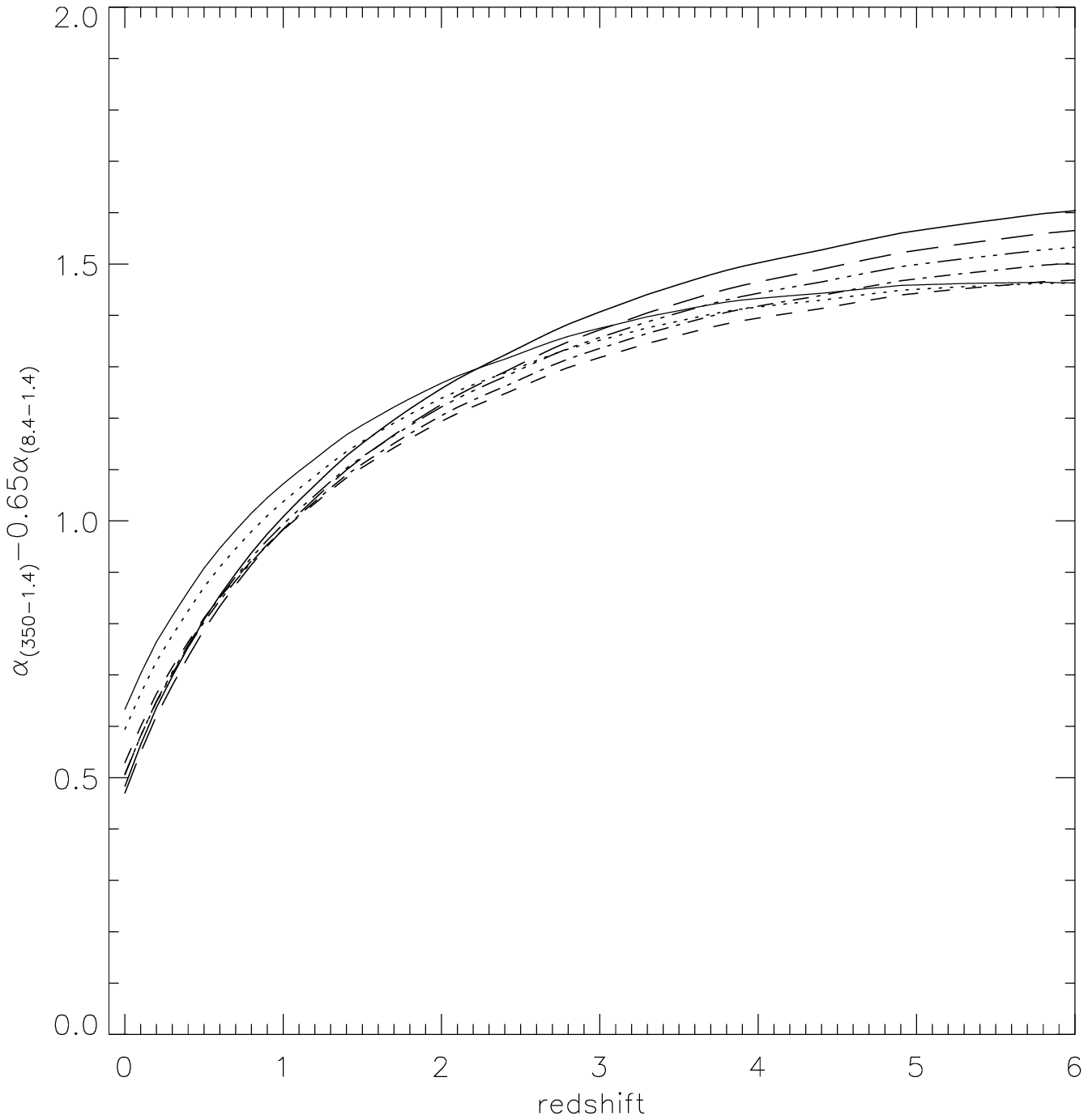, height=5truecm,width=5truecm} }        
\caption[]{{\it Left}: $850 \mu$m-$1.4$ GHz spectral index vs z for the ARP220 model observed at different ages during
the starburst. {\it Right}: linear combination of the ($850 \mu$m-$1.4$ GHz) and ($8.4$-$1.4$ GHz) spectral indexes vs z
for the same models.}
\label{qzarp}
\end{figure}

The sub-mm to radio spectral index has been recently used to estimate the redshift of sub-mm selected 
galaxies (e.g.\ Carilli \& Yun 2000). 
Exploiting our model it is possible to investigate the dependence of this index on the evolutionary
status of galaxies.
In Figure~\ref{qzarp} the $850 \mu$m-$1.4$ GHz spectral index 
is plotted vs redshift using the model fitting the SED of ARP220 (Figure~\ref{m82r}, this 
seems to be representative of the SCUBA galaxies, e.g.\ Barger et al., 1999), but for different ages of the starburt. 
The redshift dispersion around z=2 (the typical value estimated for SCUBA galaxies) is $\Delta z=1.8$. 
To reduce this dispersion, we propose to use a radio-radio spectral index (e.g.\ $8.4$-$1.4$ GHz), almost independent
of z, in combination with the previous one. Indeed, with the linear combination 
shown Figure~\ref{qzarp}, the $\Delta z$ due to the evolution becomes 0.5 around z=2. 

\section{Stellar X-ray emission from galaxies}  

\begin{figure} 
\vspace{-0.3truecm}
\centerline{ \epsfig{file=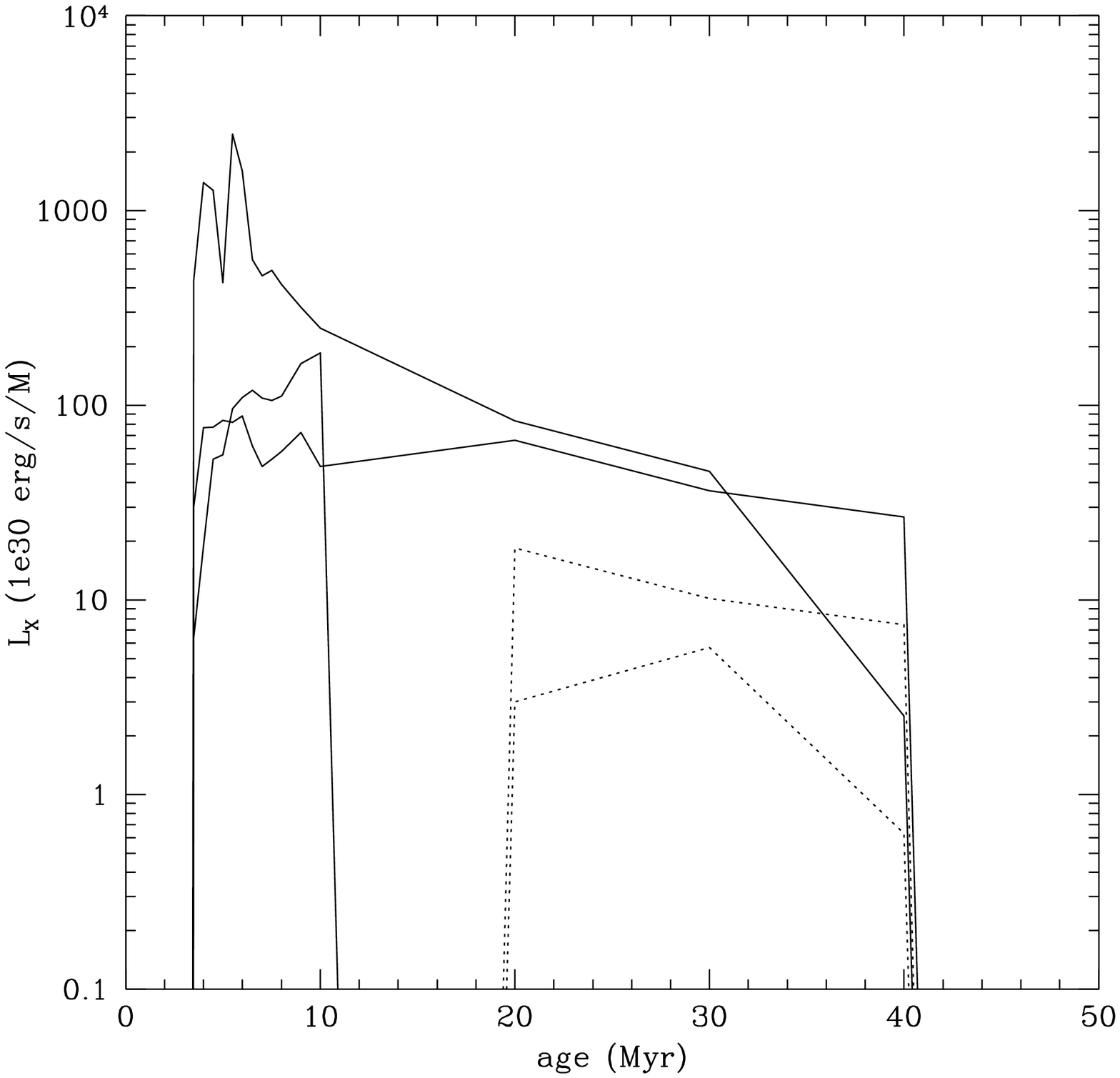, height=5truecm,width=5truecm}   
\epsfig{file=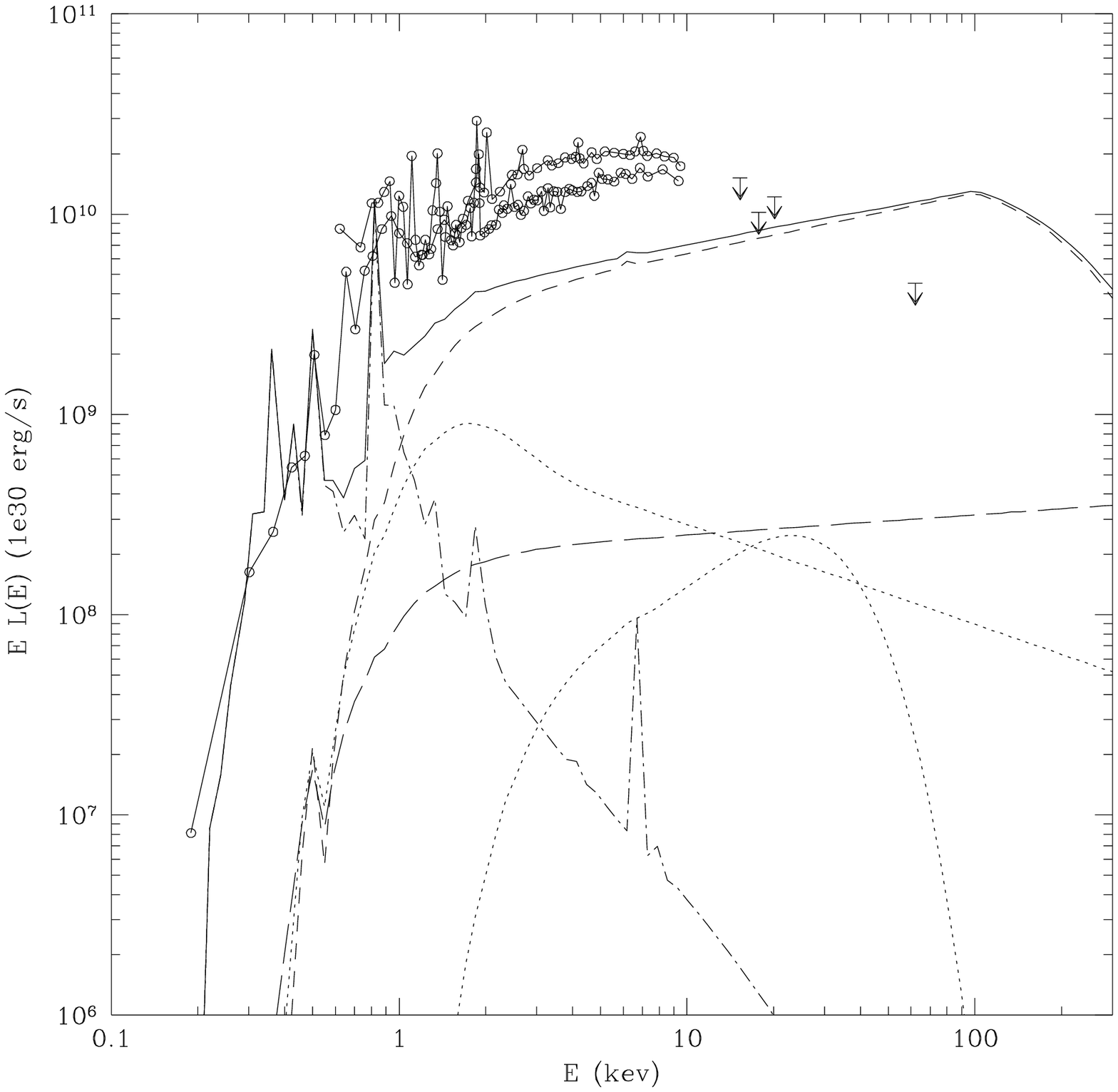, height=5truecm,width=5truecm} }          
\caption[]{{\it Left:} Predicted evolution with age of the contribution of different stellar components to the X-ray 
emission of SSPs. The minimum mass for BH formation is $14$M$_\odot$ and $P_0=0.02$ s. 
{\it Right}: Predicted stellar X-ray SED for M82 based on the UV-to-radio fitting model.
Data from Moran \& Lehnert (1997), Cappi et al.\ (1999).}
\label{sspx}
\end{figure}

Recently we have started to include in our model the stellar contribution to the X-ray emission of galaxies.
Our starting point has been the work by Van Bever \& Vanbeveren (2000). They find that 
the dominating stellar components in the X-ray 
for starburst galaxies are massive binaries with a black hole or a neutron star as a primary, and an OB star as a 
secondary either in the H-shell burning phase or on the main sequence, and pulsars.
We have included these components, and the contribution from supernova remnants, 
in the SSP library. The main uncertainties in the predicted output are (Figure~\ref{sspx}):
the binary fraction, the minimum mass for BH formation and the initial spin period $P_0$ of pulsars (whose $L_X \propto
P_0^{-2}$). A further uncertainty is the SED to assigne to each component.
A first attempt for the stellar X-ray SED expected from the UV-to-radio fitting model for M82, 
is shown in Figure~\ref{sspx}.

\end{article}
\end{document}